# Role of stochastic processes in particle charging due to photoeffect on the Moon

Eugene V. Rosenfeld[1], Alexander V. Zakharov[2 *]

[1]Institute of Metal Physics (IMP) of the Ural Branch of the Russian Academy of Sciences, Kovalevskaya str. 18, Yekaterinburg 620990, Russia. Email: evrosenfeld@gmail.com

[2*]Space Research Institute (IKI), Profsoyuznaya str. 84/32, Moscow, 117997, Russia. Email: zakharov@iki.rssi.ru, corresponding author

*Abstract*

Neglecting the effects associated with the solar wind plasma, the photoelectrons are the only elementary particles which create an electrical current through sunlit surface of the moon. They are knocked off of the surface soil, rise above the surface, and then fall back. Therefore, on average, on any unit of surface area there is a positive charge $\sigma$, equal in magnitude to the charge of photoelectrons flying over this area. However, the charge of any small dust particle can strongly fluctuate discretely: a photoelectron can be either knocked off of the particle (increase of the particle charge on $e>0$) or be reacquired by the particle (decrease of its charge on $-e$). The result is a "random walk" in sign and magnitude of the charge of grains. In a few minutes after sunrise, almost every dust particle on the surface has at least one extra or missing electron, and the average modulus of the charge accumulated on a particle is proportional to the square root of the number of "steps" (knocking off /returns of photoelectrons). Therefore, the average value of the modulus of the charge of a fine dust particle exceeds by several orders of magnitude the proportion of the average surface charge $0.25\sigma d^2$ attributable to the particle ($d$ – its diameter). So dust particles that have ejected a sufficient number of photoelectrons can take off from the surface because of the electric field of the near surface charge double layer. It is shown that: (i) almost half of all dust particles on the illuminated lunar surface are missing at least one electron; (ii) a significant portion of particles up to 100 nm in size emits several photoelectrons and acquire a positive charge, sufficient to take off from the surface; (iii) above the surface there is a "boiling layer" of dust with a maximum thickness of several hundred meters where the average size of the particles and their density non-monotonically depend on the altitude; and (iv) scattering of light in the lower part of this layer may be the cause of low-altitude horizon glow.

1. **Introduction**

The surface of the Moon, as the surfaces of other atmosphereless space objects exposed to UV radiation and streams of charged particles, is charged. In general, due to the occurrence of this charge, a balance of the fluxes of charged particles of different origin (electrons and ions of the solar wind, photoelectrons, secondary electrons and protons, etc.) is achieved. Equilibrium of the surface charge density is achieved when the perpendicular to the surface component of the total electric current vanishes [Manka, 1973]. However, we will not consider streams of charged particles falling from space to the surface and will confine ourselves to only considering photoelectrons knocked from the surface due to UV radiation. With the large radius of the Moon, its surface can be considered flat in our small study area, and in this case, all photoelectrons ejected from the surface return, so that the condition of vanishing current from the surface occurs automatically.

Indeed, the photoelectron stake off over the surface creating a layer of negative charge, a "photoelectron sheath", having a thickness of about 10 m [Stubbs et al., 2007], and thus the surface is positively charged. The result is a "capacitor", with one of the electrodes being the surface, and the other electrode is the photoelectrons layer. With a large radius of the Moon, this capacitor can be considered flat in any local area. Then the magnitude of the surface charge is equal to the charge of photoelectrons above the local surface.

Once again in this simplest approximation, we neglect the presence of other charged particles, and in this approximation the charge of a large planet (taking into account photoelectron sheath) can be considered equal to zero. The electric field $E$ arising in the condenser, according to various estimates, is about $1 Vm^{-1}$ [Freeman, Inrahim, 1975]. It is this decreasing electric field that brakes all the emitted photoelectrons and compels them to return to the surface, so the total current automatically becomes zero, and the same electric field, acting on the positively charged dust particles lying on the surface, tends to raise them.

Evaluation of particle sizes that can be lifted from the lunar surface by the action of the electric field has been implemented previously in several works (e.g., Borisov and Mall, 2006; Colwell et al., 2009). We believe, however, that the estimation of the charge of individual dust particles deduced by the authors of these works is exaggerated. Therefore, the aim of this study was to conduct a more thorough assessment of the amount of charge of dust particles lying on the surface, taking into account its fluctuations but disregarding the effects related to solar wind. This gives the possibility to estimate also the sizes of particles that can be lifted from the surface due to the electrostatic mechanism and the maximum lifting height.

## 2. The charge required to lift a dust particle

The similarity of the charged double layer to a flat capacitor occurs because the charge density of the photoelectrons at the height $h$ is inversely proportional to their speed $v(h)$ at this altitude. As a result, near the surface where the speed of taking off and returning electrons is maximum, the density of negative charge is minimal. On the contrary, in the upper part of the layer where the photoelectrons are stopped and begin to fall back, the density of negative charge is maximum. Therefore, in the simplest approximation, we can represent the double layer in the form of two plates with charge density $\pm\sigma$ separated by distance $H$ (the thickness of the photoelectron sheath). In this model, there is a very simple relations between the parameters

$$E = V/H = \sigma/\varepsilon_0 \qquad (1)$$

where $V$ is the potential difference of the plates. To establish the relationship between the V and the velocity of departure of the photoelectrons $v_0$

$$v_0 = \sqrt{2\frac{\hbar\omega - W}{m_e}} \qquad (2)$$

where we assume the radiation is monochromatic (angular frequency ω), the work function $W$ is the same for all electrons and the direction of departure is always perpendicular to the surface. Then we obtain the expression for flight time up and down $\Delta t$ and for the surface charge density σ:

$$\Delta t = 2H\sqrt{2\frac{m_e}{eV}}, \quad \sigma = j_{ph}\Delta t. \qquad (3)$$

where $m_e$ is electron mass, $j_{ph}$ is the density of the photocurrent from the surface and it takes into account that in equilibrium, at any time $t$ only those electrons suspended over the surface that were knocked out after the moment $t$-$\Delta t$.

Various authors [Goertz, 1989; Manka, 1973] agree that $j_{ph} \approx 4.5\,\mu A\,m^{-2}$ at lunar noon, i.e. about $3 \times 10^{13}$ electrons are released every second from a surface area of 1 m$^2$, and the same amount is returned back when at equilibrium. Substituting this value in (1) - (3), we obtain

$$\sigma = 2\sqrt{\varepsilon_0 j_{ph}} \left(\frac{m_e}{2e}\right)^{1/4} V^{1/4} \approx 1.5 \cdot 10^{-11} V^{1/4}, \text{ C m}^{-2};$$

$$H = \frac{1}{2}\sqrt{\frac{\varepsilon_0}{j_{ph}}} \left(\frac{2e}{m_e}\right)^{1/4} V^{3/4} \approx 0.5 \cdot V^{3/4}, \text{ m}; \quad (4)$$

$$\Delta t = 2\sqrt{\frac{\varepsilon_0}{j_{ph}}} \left(\frac{m_e}{2e}\right)^{1/4} V^{1/4} \approx 3.7 \cdot 10^{-6} V^{1/4}, \text{ s}.$$

The value of the potential difference $V$ in the photoelectrons sheath has units of volts.

Thus, when $V \approx 10$ V the average density of positive charge on the surface is of the order $\sigma \approx 3 \cdot 10^{-11}$ Кл м$^{-2}$, and to establish this value, i.e. for the formation of a double layer, requires about $\sigma/j_{ph} \approx 7$ μs. At the beginning of this process the time of flight of the photoelectrons is greater than $\Delta t$ (4), so the real time of the double layer stabilization will be slightly more. After this time the average charge of dust particles with diameter $d$ lying on the surface (i.e., the surface positive charge of attributable to its share) becomes and remains equal to

$$\langle q(d) \rangle = \frac{1}{4}\pi d^2 \sigma \approx 1.5 \cdot 10^8 d^2 e. \quad (5)$$

where $d$ is in meters. From this expression it is clear that $\langle q \rangle$ may exceed the elementary charge $e$ only if the diameter of a dust particle is not less than 80 microns. For smaller particles the probability to acquire this smallest possible charge falls sharply with decreasing diameter. On average only one of 650,000 grains with a diameter of $d=100$ nm and one of 65,000,000 with diameter $d=10$ nm has a charge of $e$.

It is easy to find the number of electrons $n_0$, which a dust particle with a diameter $d$ needs to lose, in order for the field of strength $E$ to be able to pick it up:

$$n_0(d) \approx \frac{\rho g d^3}{2eE} \approx 1.5 \cdot 10^{22} \frac{d^3}{E}, \quad (6)$$

where $g \approx 1.6$ m s$^{-2}$ is the acceleration of free fall on the Moon, $\rho \approx 3000$ кг м$^{-3}$ is the density of the lunar soil [Vaniman et al., 1991]. Thus, for $E \approx 1$ V m$^{-1}$ the one electron charge is sufficient to raise a dust particle with a diameter less than 40 nm, but a dust particle 100 nm in diameter will need a charge of $15e$, and for micron size particle - $15000e$ etc. Therefore it might seem, that the electric field in the double layer is able to raise only a tiny fraction of the smallest grains.

## 3. "Random walk" of the charge of dust particles

In fact, however, the situation is not so simple, because the electrons are not only emitted from the surface, but fall back just as often. This means that during the time

$$\delta t(d) = \frac{4e}{\pi d^2 j_{ph}} \approx 4 \cdot \left(\frac{10^{-7}}{d}\right)^2, \text{ s} \qquad (7)$$

on average, one electron impacts a dust particle with a diameter of *d* m (or is knocked out from it). According to this formula, one "step", i.e. change of charge by $\pm e$ takes about 4 sec for dust particles with diameter 100 nm and about 6 - 7 minutes if its diameter is 10 nm. As a result of these "random walks" after a few minutes after the occurrence of the photocurrent the charge of one sign or another should appear on almost all particles with a diameter of 10 nm or more. The average number of missing or extra electrons on the individual dust particles is proportional to the square root of "number of steps" [see e.g. Van Kampen, 1992], but on the surface the number of particles with an excess of electrons is only slightly less than with a deficit, and their charges in large part compensate for each other. As a result, the average surface charge, and with it the average charge of dust particles (5) is very low.

Since the "charging time" of the dust particles $t = \delta t(d) \times (\text{number of steps})$ is proportional to the number of "steps", we can say that the average value of the modulus of the charge increases in proportion to $\sqrt{t}$. It is important, however, to keep in mind that the duration of the step (7) depends strongly on the size of the dust particles. Denoting by $q(d,t)$ a charge of a dust particle accumulated by the time *t*, we get:

$$n(d,t) \equiv \frac{q(d,t)}{e}, \quad |n(d,t)|_{most\ prob} \approx \sqrt{\frac{t}{\delta t(d)}}. \qquad (8)$$

Here, we put the sign of approximate equality in order to emphasize that although the charge of dust particles may be equal only to an integer number of elementary charges, we consider *n* as the smoothed value (i.e. averaged over intervals on which a lot of "steps" are placed), and therefore a continuous function of time.

Slow but unlimited growth of the average values of the modulus of the charge described by this formula in reality will only occur if the probabilities of knocking out an electron from a dust particles and falling back onto the particle are the same. In this case we come to the task reminiscent of the classical "Gambler's Ruin Problem" [Epstein, 1995]. However, this analogy is valid only for neutral dust particles. In the case of a particle or a neighboring particle acquires a charge the situation is changed. For positively charged dust particles, the probability of an electron to fall on a dust particle must be higher than the probability of fly off, and for negatively charged particles the opposite would be true. Using the terminology of "the Gambler's Ruin Problem", the principal features of the emerging problem now can be formulated as follows: (i) while increasing a gambling gain/loss of the player his original symmetric coin (the probability of heads and tails is the same) becomes more and more asymmetric; (ii) the sign of the asymmetry is always such to reduce the gain/loss.

To solve such problem accurately is less likely especially because it is not known how the probability of an electron hit on a dust particle depends on its charge. Nevertheless, it is clear the general feature of the required the time dependence of a modulus of the charge: initially it has to grow proportionally to $\sqrt{t}$, but while the probability of an electron falling on a dust particle changes, the rate of change of the charge is decreasing. As a result, the charge modulus should reach a maximum and then decrease (possibly oscillating) to a sign change and start a new cycle. We are ultimately only interested in how the value of the maximum charge (which determines the maximum lifting force) and the time of its accumulation depend on the size of dust particles. Below we look to answer both questions.

**4. The influence of the charge of a dust particle on the rate of its change**

We can try to find a simulacrum of the solution to this problem, reducing it to a differential equation that determines the "smoothed" rate of the charge magnitude change of dust particles $ne$:

$$\frac{dn}{dt} = \frac{\text{sign}(n)}{2\sqrt{\delta t(d) \cdot t}} + \frac{\tau(d,n) - \delta t(d)}{\delta t(d) \tau(d,n)}, \quad n(t=0) = 0. \tag{9}$$

The first term in the right part of (9) describes increasing of the modulus $n(t)$ (the derivative of this function $\sqrt{t}$ is inversely proportional to the root) due to random processes (8), the second term defines the linear in time reduction $|n(t)|$ associated with the difference in the mean time intervals between the departure of photoelectrons from a dust particle $\delta t(d)$ (7) and their hits on the particle $\tau(d,n)$. Suppose in the simplest approximation, that

$$\tau(d,n) \approx \delta t(d)(1 - \alpha n) \tag{10}$$

Where $\alpha$ is a dimensionless parameter, which determines how strongly the probability of electrons hit on a dust particle changes with its charge $ne$. Now, assuming α is small and limited in (9) only by linear in $n$ terms, we obtain the equation:

$$\frac{dn}{dt} \approx \frac{\text{sign}(n)}{2\sqrt{\delta t(d) \cdot t}} - \alpha \frac{n}{\delta t(d)} \quad . \tag{11}$$

The solution of this equation is the function $n(d,t)$:

$$n(d,t) = \frac{\pm 1}{\sqrt{\alpha}} f\left(\frac{t}{t_0}\right), \quad t_0(d) = \frac{\delta t(d)}{\alpha} \Rightarrow \frac{t}{t_0} = \alpha \frac{t}{\delta t(d)};$$
$$f(x) = \exp(-x) \int_0^{\sqrt{x}} \exp(u^2) du. \tag{12}$$

Introduced here function $f(x)$ that does not depend on $\alpha$ and on $d$ initially increases rapidly with increasing $x$ from zero, reaches the maximum value $f_0 \approx 0.54$ in the point $x_0 \approx .84$ and then slowly decreases to zero as $\frac{1}{2\sqrt{x}}$ (see the description of the function at the end of the preceding chapter). This means that on average the charge amount of dust particles first increases during time $t_{max} \approx x_0 \alpha^{-1} \delta t(d)$ seconds up to the value $|q_{max}| \approx f_0 e \alpha^{-\frac{1}{2}}$ and then slowly decreases until the beginning of a new cycle (see Fig. 1). For example, if $\alpha = 0.001$ the average value of the modulus of the charge of dust particles reaches its maximum $f_0 \alpha^{-\frac{1}{2}} \approx 17$ elementary charges after about $x_0 \alpha^{-1} \approx 840$ steps (photoelectrons emission / absorption). This process of "charging" will last about an hour for dust particles with a size of 100 nm ($\delta t(100 \text{ nm}) \approx 4$ s, see (7)) and about four earth days for dust particles with a size 10 nm ($\delta t(10 \text{ nm}) \approx 400$ s). Thus, only the value $\alpha$ determines the maximum (in average) number of excess / missing electrons that can be accumulated on the dust particles, and the average number periods of time $\delta t(d)$, that must pass away until the maximum charge accumulation. The real duration of this process (with the duration of "steps" $\delta t(d)$) varies depending on the dust particle size. In addition, it should be noted that the formula (12) and the graphs in Fig. 1 describe only the average charge modulus change with increasing of number of "steps" (although, as in the toss of a coin, there may be significant, although rare, deviations from the mean).

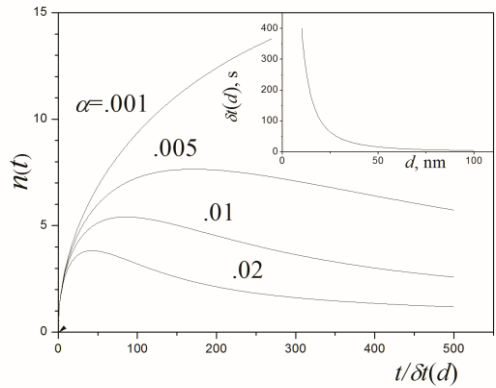

Figure 1. Average number of accumulated electronic charge $n(\alpha, t)$ (12) on the time (indeed the x-axis is the number of "steps" with a duration of $\delta t(d)$ (7), see the note in the chapter 3). Insertion: Dependence of the "step" duration $\delta t(d)$ (the time between photoelectrons escapes from a dust particle) on its size.

**5. The maximum size of a flying up dust particle**

Using the approximation (10) postulates that the difference between $\delta t(d)$ and $\tau(d,n)$ is proportional to the particle charge. This is quite natural in the case of a solitary dust particle. More precisely, in this case, instead of the charge $q$ we need to use the potential of particle surface proportional to $q/d$. Falling electrons drawn into the areas with high potential and pushed away from the areas with low potential, will more frequently fall on a positively charged dust particle, and scarcer in the negative charged particle. In our case, however, the small dust particle lies on the surface with a plurality of other particles, each of which carries its own charge, and generates its own field. Therefore, the problem arises of how changing the probability of the electrons hit into the central particle due to the appearance or redistribution of charges on the dust particles that surround it. To solve this problem exactly is unlikely. To simplify this problem we use a formula for a solitary dust particle and introduce a parameter $\alpha$, which describes all the other factors affecting the change in the probability $w \equiv \tau^{-1}$ of an electron hit on a dust particle. This is the meaning of the formula (10).

However, it is hoped that the problem to realistically assess the magnitude $\alpha$ can still be solved. On the one hand, it is possible, apparently, to try to set up a laboratory experiment to determine the charge magnitude of dust particles on a surface, illuminated with ultraviolet light. On the other hand, perhaps some progress could be made in the development of methods to calculate the neighboring particle charges that have been applied in [Vaverka et al., 2016, Wang et al., 2016]. In any case, however, it should be noted that as the maximum charge of dust particles $|q_{max}| \approx f_0 e \alpha^{-1/2}$ is inversely proportional only to the square root of $\alpha$, the magnitude of this charge (and with it the maximum amount of the lifting force) varies with a change of $\alpha$ slowly.

In order to estimate the value $\alpha$, assume that the change of $\tau$ is proportion to the ratio of the potential energy of the electron on the dust particle surface to the kinetic energy that the incident electron had when approaching it:

$$\frac{\delta\tau(n)}{\tau(n=0)} \approx -\frac{ne^2}{4\pi\varepsilon_0 \tfrac{1}{2}d \tfrac{1}{2}m_e v_0^2} \approx -\frac{2ne}{4\pi\varepsilon_0 Vd} \Rightarrow \alpha \approx -\frac{3\cdot 10^{-9}}{Vd} \quad (13)$$

($V$ in volts, a particle diameter $d$ in meters; compare this formula with (10)). It is considered that the kinetic energy of the incoming photoelectrons is determined by the potential of a double layer $m_e v_0^2 = 2eV$, and in this approximation $\alpha$ is dependent on $d$. At $V = 10$ V we get $\alpha \approx 0.03$ for particles with a diameter $d = 10$ nm and $\alpha \approx 0.003$ for $d = 100$ nm. It is natural to expect, however, that the presence of random fields generated by neighboring particles should smooth out the effect of the central field of dust particles, reducing the value of $\alpha$.

As seen from Fig. 1, the magnitude of $|q_{max}|$ is on the order of ten elementary charges (range 4-15) in a fairly wide range of values $\alpha$ from 0.001 to 0.020. Since the maximum possible size of the flying dust particles is changing more slowly (proportional to the cubic root of the lifting force (6)), the charge $|q_{max}| \sim 10e$ will be able to raise a dust particle, the diameter of which exceeds not more than two or three times the diameter of the dust liftable by the charge $e$ $(d \approx 40 \text{ nm})$. For example, with increasing the diameter of dust particles by three times, its charge and the lifting force is increased 27 times.

### 6. "Boiling layer" of dust over the lunar surface

A dust particle, in order to take off, must accumulate a charge of $ne$, so the Coulomb force $F_Q = neE$ acting on it becomes more than its weight $F_g = \tfrac{\pi}{6}\rho g d^3$ (without cohesive forces). Because the charge changes discretely, a dust particle should emit $(n_{max} - 1)$ electrons to not fly off, and only after the loss of the last, $n_{max}$-th electron does the Coulomb force exceeds the gravity force. Therefore, the total force (sum of Coulomb`s and gravitational forces $F = F_Q - F_g$), which raises a speck of dust from the surface never exceeds the force $-F_g \leq F \leq eE$ acting on one electron. Denote by

$$\delta q = F/E \leq e \quad (14)$$

The amount of "excess" charge, which boosts a dust particle moving inside of the photoelectron sheath (the force acting on the rest of the charge $n_{max}e - \delta q$, just compensates for the weight). Accordingly, the kinetic energy, which the dust will receive while passing all of the double layer, is equal to $V\delta q$, so it is able to rise to its maximum height

$$h_{max} \approx H + \frac{6\delta q V}{\pi \rho g d^3} \approx H + 0.07 V \frac{\delta q}{e} \left(10^{-7}/d\right)^3 \text{ m} \quad (15)$$

($d$ in meters, $V$ in volts). Dust particles that are the lightest "in their $n$-th weight category" (i.e. the lightest among the dust particles that are still lying on the surface having a charge $(n-1)e$, but taking off after emitting only one additional photoelectron) reach the maximum height. For these particles $F_Q$ is almost compensated the weight before the emission of the last electron, so $\delta q$ for them is maximal, and they have less weight.

In the formula (15) it is assumed that all dust particles have the same density and are spheres. If the potential difference in the double layer near the surface is $V \approx 10$V, $H \approx 10$ m and $E \approx 1$ Vm$^{-1}$ so the Coulomb force acting on a charge of $1e$ balances the weight of dust particles with diameter $D_1$=40 nm, we can also write

$$D_n = D_1 n^{1/3}, \quad \delta q(d) = e \frac{D_{n_{max}}^3(d) - d^3}{D_1^3} \quad (16)$$

where $D_n \approx$ 40, 50.4, 57.9, 63.5, ...nm - diameters of the dust particles, the weight of which is balanced by the Coulomb force acting on charge $ne$ ($n$=1,2,3,4,...). $D_{n_{max}}(d)$ - the maximum diameter of the dust particles, which, like a dust particle with diameter $d$, takes off, having a charge of $n_{max}e$. The dependence of the maximum height (15) and the excess of the charge (14) of the particles on their diameter are presented in Fig. 2.

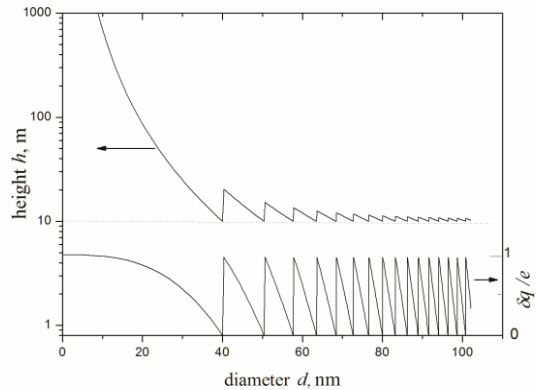

**Figure 2**. The dependence of the maximum height of the particle (upper curve, left logarithmic scale) and the excess charge (lower curve, right scale) on the diameter of the dust particles. Dashed line represents the position of the upper boundary of the double layer.

From (15) and Fig. 2 it is clear that a dust particle with diameter of 10 nm (it is almost ready to take off because gravity acting on it tens times less then $eE$) after the emission of a photoelectron will rise to a height of about 700 m. With increasing diameter that increases the mass, and though one electron charge is still sufficient to raise a dust particle, the maximum height of their flying decreases. However, once the diameter of the dust particles exceeds 40 nm, it would have to emit two photoelectrons to leave the surface. Almost all the Coulomb force acting on the second electron within the double layer will be used for its acceleration, as for dust particles with $d$=10 nm. However, for 40 nm the mass is 64 times larger, and the height above the upper boundary of the double layer is only about 12 m and this height decreases with increasing of the particle diameter.

We see that the discrete nature of the change of the dust particles charge leads to the formation of a "boiling layer" of charged dust particles on the illuminated surface of the Moon. However, the density distribution of dust in this layer is very radically different from the distribution of photoelectrons. At the surface, where the departing dust particles have not had enough time to acquire sufficient speed and the returning particles completely lose their vertical velocity, the speed of these particles is minimum and their density maximum (a fundamental difference from the density of photoelectrons!). With the increase in height, the dust`s speed increases, reaching a maximum at the height $H$ equal to the thickness of the double layer, and at this altitude the density of the particles is minimal. Further with height the speed decreases again, but for different particle sizes the maximum lift height is different. Therefore there should occur a sparser layer of dust than near the surface, where particles of all sizes have small speed.

In this idealized picture, we should make some corrections.

Firstly because of the spreading of the values and directions of velocities of photoelectrons departing from the surface the double charge layer above the surface has no clear upper bound. This means that the field strength varies with the altitude and the region in which particle velocity reaches its maximum, is blurred. Accordingly, the dependence of the particle density on the height is blurred too.

Secondly, each cycle of the dust particle motion is a parabola (rise-fall) and taking into account that the free fall acceleration is relatively small, the cycle goes on long enough, so there is a finite probability, that during the flight, the dust particle can absorb a photon and emit one more photoelectron. In this case, the returning dust particle will not reach the surface, and stops and starts again to fly somewhere within the double charge layer, which will lead to even greater blurring of the distribution function of particles over the surface of the planet. On the other hand, inside the double layer the probability that the dust particle will absorb an electron increases (here electrons approach to a dust particle from above and from below), so its charge will decrease and it will fall to the surface.

Thirdly, it should be emphasized that small uncharged dust particles also must move up and down within thin layer above a hot surface due to thermal fluctuations [Rosenfeld et al., 2016].

As a result of all these processes the calculation of the distribution function of dust particles as a function of sizes and heights becomes a very difficult task. However, it is clear that the layer of "dancing" particles should be most dense near the surface and became more thin (probably non-monotonic), to extend up to heights somewhat greater thickness of the double layer. Further, up to the limiting height of several hundred meters, only the smallest particles can occur. It is natural to assume that the light scattering in the lower, most dense part of this layer is associated with the emergence of low altitude horizon glow [Criswell, 1973; Rennilson, Criswell, 1974; Severny et al., 1975]. However, dust particles at the high altitude [Zook et al., 1995, Wooden et al., 2016] and streamers with an altitude of hundreds of kilometers [McCoy, J.E., Criswell, D.R., 1974] can't be associated with this layer.

7. **Conclusion**

The process of accumulating charge on dust particles, lying on the illuminated surface of any atmosphereless celestial body, has two principal features. First, it is a quantum process, and, secondly, also a stochastic process. The charge of dust particles can change only discretely by $\pm e$ as a result of the random emission of an own photoelectron or absorption an electron, emitted earlier from another dust particle. Therefore, there is "random walk" of charge of any dust particle, the amplitude of which is much higher than the value of the share of the average surface charge that is attributable to it. The probability of an electron hitting a dust particle depends on the sign and magnitude of the charge of this particle and the sign and magnitude of the charges of neighboring grains. It's clear that this effect leads to the suppression of large charge fluctuations but we could only highly approximately describe the process. For another thing, the question on attraction between oppositely charged neighboring grains that encumbers the taking off of a single particle remains a challenging open problem.

Analyzing the effects caused by these "random walks" using the simplest models we arrive at the following semi-quantitative conclusions:

1. The time from the start of solar lighting to the stabilization of the average value of the surface charge, the formation of the double layer of photoelectrons and the average electric field above the surface is about a few microseconds (see text after (4).). The density of the surface charge $\sigma$ is small, so the average value of the charge of fine dust particles (5) is several orders of magnitude less than the charge of an electron $e$. Only the average charge of the dust particles with a diameter $d > 80$ μm is greater than $e$.

2. A few minutes after the onset of the photocurrent density $j_{ph} \approx 4.5\,\mu A\,м^{-2}$, surplus or lack of at least one electron arises on practically every dust particle. The dependence of this time on the size of dust particles is determined by the formula (7), see also the inset to Fig.1. At a later stage, the particle charge fluctuates with the greater amplitude, the larger diameter. Although the charge of every separate dust grain is large enough, only a little more than half of charged dust particles are positively charged, the other have negative charge. This is why their average charge is low (see previous point).

3. With a lack of a single electron, only dust particles with diameters of not more than $d \approx 40$ nm are able to take off from the surface of the moon in the field of $E \approx 1\,V\,m^{-1}$. A grain with a diameter $k$ times larger (having a mass $k^3$ time greater), would require $k^3$ more missing electrons to be lifted from the surface. In a wide range of parameters, the average maximum value of the modulus of dust particle charge is on the order of $10e$. This means that the particles with sizes up to 100 nm or slightly larger can rise over the surface in a few hours after the beginning of lighting in the field $E \approx 1\,V\,m^{-1}$.

4. The lifting height of particles above the surface depends on their size (see formula (15) and the text following the formula), but for particles with $d$>10 nm this height may not exceed several hundred meters. The result is a "boiling layer" of particles of different sizes, which may be the cause of the low-altitude horizon glow observed directly above the surface of the Moon by the Surveyor landing modules [Rennilson, Criswell, 1974].

**Acknowledgement:**

This work was supported in part by the state assignment of FASO of Russia, theme «Kvant» № 01201463332 (IMP) and by the program I.7 of the Russian Academy of Sciences "Experimental and theoretical research of Solar system objects and planetary system of stars" (IKI).
The authors gratitude Prof. Thomas Duxbury for help in preparation of this paper.

**References**

Criswell D. R., 1973, Horizon-glow and the motion of lunar dust, in: Photon and Particle Interactions With Surfaces in Space, edited by R. J. L. Grard, pp. 545–556, Springer, New York.

Epstein K., *The Theory of Gambling and Statistical Logic,* Elsevier, 1995 (ISBN: 978-0-12-374940-6)

Freeman J.W., Ibrahim M., 1975, Lunar electric fields, surface potential and associated plasma sheaths, The Moon 14, pp. 103-114

Goertz C.K., 1989, Dusty plasma in the solar system, Rev. Geophys., 27, pp. 271-292.

Manka R.H., 1973, Plasma and potential at the lunar surface. in: R.J.L. Grard (Ed.), Photon and particle interactions with surfaces in space. D. Reidel Publishing Co., Dordrecht, Holland, pp. 347-361.

McCoy J. E., and D. R. Criswell, 1974, Evidence for a high latitude distribution of lunar dust, Proc. Lunar Sci. Conf., 5th, 2991.

Rennilson J.J., Criswell, D.R., 1974. Surveyor observations of lunar horizon-glow. Moon 10, 121–142.

Rosenfeld E.V., Korolev A.V., Zakharov A.V., 2016, Lunar nanodust: Is it a borderland between powder and gas? Advances in Space Research 58, 560–563.

Severny A. B., E. I. Terez, and A. M. Zvereva, 1975, The measurements of sky brightness on Lunokhod-2, Moon, 14, 123–128.

Stubbs T, R.R. Vondrak, W.M. Farrell, 2006, A dynamic fountain model for lunar dust. *Adv. Space Res.* **37**, pp. 59-66.

Van Kampen N. G., 1992, Stochastic Processes in Physics and Chemistry, revised and enlarged edition (North-Holland, Amsterdam), and refs. therein.

Vaniman B., R. Reedy, G. Heiken, G. Olhoeft, W. Mendell, 1991, The Lunar environment, in: *Lunar Sourcebook*, G.H. Heiken, D.T. Vaniman, B.M. French, Eds. (*Cambridge Univ. Press, 1991)* p. 27. (http://www.lpi.usra.edu/publications/books/lunar_sourcebook/).

Vaverka J., Richterová I., Pavlů J., Šafránková J., Němeček Z., 2016, Lunar surface and dust grains potentials during the Earth`s magnetosphere crossing, Astrophys. J. vol. 825, No. 2.

Walbridge E., 1973, Lunar photoelectron layer, J. Geophys. Res., vol. 78, No. 19, pp. 3668-3687.

Wang X., Schwan J., Hsu H. –W., Grün E., and Horányi M., 2016, Dust charging and transport on airless planetary bodies, Geophysical Research Letters, June (DOI: 10.1002/2016GL069491)

Wooden D. H., Cook A. M., Colaprete A.D., Glenar A., Stubbs T. J. and Shirley M., 2016, Evidence for a dynamic nanodust cloud enveloping the Moon, Nature Geoscience, DOI: 10.1038/NGEO2779.

Zook H. A., A. E. Potter, and B. L. Cooper (1995), The lunar dust exosphere and Clementine lunar horizon glow, Lunar Planet. Sci., 26, 1577–1578.